\documentclass[10pt,twocolumn,letterpaper]{article}

\usepackage{wacv}
\usepackage{times}
\usepackage{epsfig}
\usepackage{graphicx}
\usepackage{amsmath}
\usepackage{amssymb}
\usepackage{booktabs}

\usepackage{multirow}
\usepackage{tabularx}
\usepackage[accsupp]{axessibility}  

\def\wacvPaperID{1330} 

\wacvapplicationstrack 

\wacvfinalcopy 

\ifwacvfinal
\usepackage[breaklinks=true,bookmarks=false]{hyperref}
\else
\usepackage[pagebackref=true,breaklinks=true,colorlinks,bookmarks=false]{hyperref}
\fi

\renewcommand\sectionautorefname{Sec. }
\renewcommand\equationautorefname{Eq. }
\renewcommand\tableautorefname{Tab. }
\renewcommand\figureautorefname{Fig. }

\makeatletter
\newcommand*{\rom}[1]{\expandafter\@slowromancap\romannumeral #1@}
\makeatother

\begin{document}


\title{Ordinal Classification with Distance Regularization \\ for Robust Brain Age Prediction}

\author{
\vspace{5pt}
Jay Shah\textsuperscript{1,2}, Md Mahfuzur Rahman Siddiquee\textsuperscript{1,2}, Yi Su\textsuperscript{2,3}, Teresa Wu\textsuperscript{1,2}, Baoxin Li\textsuperscript{1,2}\\
\textsuperscript{1}Arizona State University, \textsuperscript{2}ASU-Mayo Center for Innovative Imaging, \\
\textsuperscript{3}Banner Alzheimer’s Institute
}


\maketitle
\thispagestyle{empty}

\begin{abstract}

Age is one of the major known risk factors for Alzheimer's Disease (AD). Detecting AD early is crucial for effective treatment and preventing irreversible brain damage. Brain age, a measure derived from brain imaging reflecting structural changes due to aging, may have the potential to identify AD onset, assess disease risk, and plan targeted interventions. Deep learning-based regression techniques to predict brain age from magnetic resonance imaging (MRI) scans have shown great accuracy recently. However, these methods are subject to an inherent regression to the mean effect, which causes a systematic bias resulting in an overestimation of brain age in young subjects and underestimation in old subjects. This weakens the reliability of predicted brain age as a valid biomarker for downstream clinical applications. Here, we reformulate the brain age prediction task from regression to classification to address the issue of systematic bias. Recognizing the importance of preserving ordinal information from ages to understand aging trajectory and monitor aging longitudinally, we propose a novel ORdinal Distance Encoded Regularization (ORDER) loss that incorporates the order of age labels, enhancing the model's ability to capture age-related patterns. Extensive experiments and ablation studies demonstrate that this framework reduces systematic bias, outperforms state-of-art methods by statistically significant margins, and can better capture subtle differences between clinical groups in an independent AD dataset. Our implementation is publicly available at \href{https://github.com/jaygshah/Robust-Brain-Age-Prediction}{https://github.com/jaygshah/Robust-Brain-Age-Prediction}.

\end{abstract}

\begin{figure}[!htp]
    \centering
    \includegraphics[trim={0.3cm 0.4cm 0 0.4cm},clip,width=\linewidth]{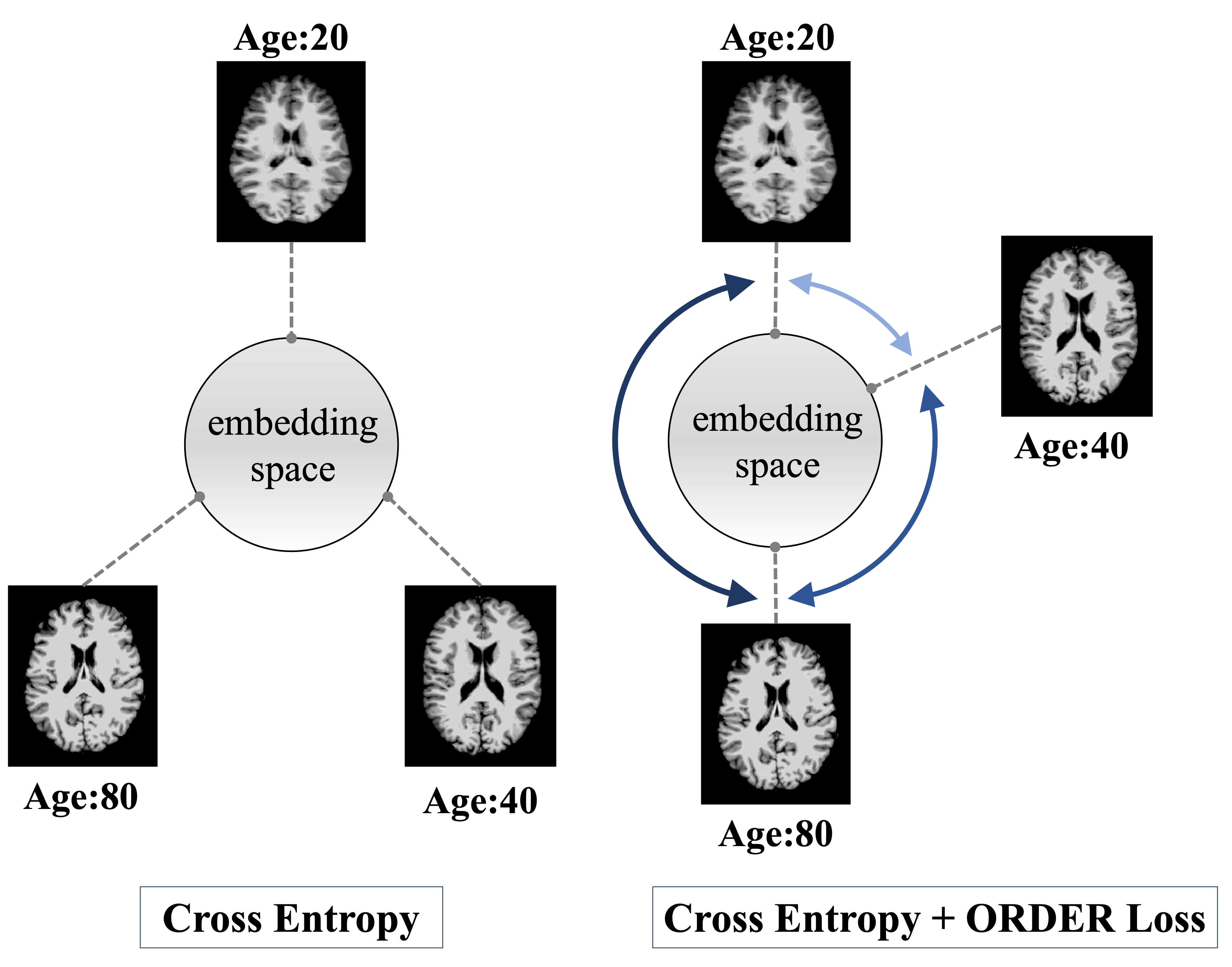}
    \caption{Standard cross-entropy vs. cross-entropy with ORDER loss: Cross entropy loss (left) encourages the model to learn high entropy feature representations where embeddings are spread out. However, it fails to capture ordinal information from labels. Our proposed ORDER loss with cross entropy (right, \equationautorefname\ref{eq:our_loss_total}) preserves ordinality by spreading the features proportional to Manhattan distance between normalized features weighted by absolute age difference. The illustrated example (right) shows embedding space where learned representations of MRI scans with ages $20$, $40$, and $80$ are distributed apart from one another, with distances proportional to absolute age differences.}
    \label{fig:order_loss_figure}
\end{figure}

\vspace*{-5mm}
\section{Introduction}
\label{sec:introduction}

Normal aging causes structural changes in the human brain across the adult lifespan, a major risk factor for the decline in physical health and cognitive ability \cite{cole2018brain}. Aging also exposes an individual to an increased risk of cancer \cite{lopez2013hallmarks} and various neurological disorders such as Parkinson's Disease \cite{beheshti2020t1}, vascular dementia \cite{haan2004can}, mild cognitive impairment (MCI) and Alzheimer's Disease (AD) \cite{franke2012longitudinal}. However, aging in humans is a complex and heterogeneous phenomenon. Even though each individual ages at the same rate chronologically, their biological age does not follow the same trajectory due to genetic factors, environmental influences, underlying neurological conditions, and other unknown factors \cite{lopez2013hallmarks}. Measuring this deviation from normal aging can allow a better understanding of associations between cognitive impairment and aging \cite{draper2008cognitive, gaser2013brainage} and identify patients at risk for clinical trials \cite{cole2018brain}. Hence, there is a growing interest in predicting biological age, most commonly derived from an individual's structural MRI data. The difference between predicted biological age and chronological age, also known as Brain Age Gap Estimate (BrainAGE) \cite{franke2010estimating}, can be used to monitor accelerated or decelerated brain aging. 

Measuring deviation from normal aging relies heavily on the base model's performance to predict normal aging, i.e., accurately predicting the biological age of healthy subjects. A model's performance on a healthy cohort is often assessed using mean absolute error (MAE), which calculates the mean of absolute BrainAGE. Existing deep learning-based regression approaches \cite{jonsson2019brain, cole2017predicting, ito2018performance} have limited clinical applications because the models have reported MAE $4-5$ years in healthy cohorts, suggesting the lack of discriminative power to interrogate BrainAGEs of different clinical groups \cite{gaser2013brainage}. Moreover, a common challenge in brain prediction models is the issue of systematic bias \cite{beheshti2019bias, liang2019investigating, zhang2023age}, where there is an overestimation in the predicted age of young subjects and an underestimation of old subjects. If BrainAGE were to be used as a reliable imaging biomarker for measuring brain health, the effect of systematic bias is of concern. For example, Alzheimer's disease patients are often of age 50+, and age underestimates will impact early detection. Studies have investigated whether this bias is induced due to model or data selection used for training \cite{liang2019investigating}. We argue that systematic bias is inherent to brain age prediction due to its formulation as regression analysis. This study has two primary objectives: (1) addressing systematic bias to enhance the robustness of brain age prediction and (2) enhancing the model's performance in predicting normal aging in healthy cohorts, thereby facilitating more accurate disease detection in downstream tasks.

Traditionally, brain age estimation is formulated as a regression task since the problem of interest is understanding which bio-signatures from imaging data have a statistically significant effect on age. More importantly, it is clinically relevant to study how these signatures change across different age groups and track their progression. To accomplish this, capturing ordinal information from the ground-truth age is critical; hence, regression is preferable. However, it is known that regression models suffer from systematic bias. To address this issue, we propose reformulating the task of brain age prediction as a multi-class classification. However, in classification, each class is treated independently of the other and hence cannot capture the ordinality of target labels \cite{zhang2023improving}. To counter this, we propose a novel {\em ORdinal Distance Encoded Regularization} (ORDER) loss in conjunction with cross-entropy loss for multi-class ordinal classification. ORDER loss is calculated based on the Manhattan distance between samples in the training mini-batch within both feature space and target space. As depicted in \figureautorefname{\ref{fig:order_loss_figure}}, it scales the distance between learned features in high-dimensional space by a weighted magnitude of the chronological age difference (see~\sectionautorefname~\ref{sec:ordinal_distance_regularization_loss}). A new ordinality metric is proposed here to quantify the relative ordering of feature representations compared to their actual target label ordering. Results show that our proposed framework preserves ordinality in feature space and improves brain age prediction by a statistically significant amount compared to existing deep learning approaches \cite{jonsson2019brain, cole2017predicting, ito2018performance}. 

One challenge in medical imaging is heterogeneity in the quality of MRI scans due to different scanners and acquisition protocols. Several studies have confined themselves to a single cohort to train and evaluate model performance \cite{jonsson2019brain}, which could affect multi-site studies or generalization performance. Contrary to that, it is shown that deep learning \cite{maartensson2020reliability} and machine learning models \cite{franke2010estimating, franke2012longitudinal, kaufmann2019common} are not only robust to scanner differences, but diversity in data due to heterogeneous sources can improve model generalization. In this study, we decide to combine cohorts from 5 public data sources to train and validate our model collected from $(1)$ National Alzheimer’s Coordinating Center's (NACC), $(2)$ Open Access Series of Imaging Studies (OASIS) \cite{marcus2007open, marcus2010open}, $(3)$ International Consortium for Brain Mapping (ICBM), $(4)$ Information eXtraction from Images (IXI), and $(5)$ Autism Brain Imaging Data Exchange-\rom{1} (ABIDE) \cite{di2014autism}. Additionally, disease detection performance is evaluated on an independent dataset. In summary, the main contributions of this research are the following:

\begin{enumerate}
    \item We formulate Brain Age prediction as an ordinal classification task that outperforms existing regression-based methods by a significant margin.
    \item A novel {\em ORDER loss} is introduced for classification that preserves the ordinality in the learned feature space from target labels, which here is Age.
    \item Proposed framework addresses the well-observed issue of systematic bias in predicted biological age from neuroimaging data.
    \item Developed model detected subtle differences between clinical groups of Alzheimer's disease, which were not accurately captured by the regression model or other approaches. 
\end{enumerate}

\section{Related Work}
\label{sec:related_work}

\subsection{Neuroimaging based Brain Age prediction}
\label{sec:brainagepred_literature}

Prior studies on brain age prediction from neuroimaging data \cite{franke2010estimating, cole2018brain, gaser2013brainage, valizadeh2017age, baecker2021brain, liem2017predicting, cole2015prediction, franke2012longitudinal, beheshti2020t1} use regression techniques such as gaussian process regression, support vector regression, and relevance vector regression. These approaches involve extensive pre-processing of raw structural MRI data and extracting imaging features such as cortical thickness, regional volumes, or surface area using tools such as FreeSurfer or Statistical Parametric Mapping (SPM). Input to the machine learning models are these pre-processed brain morphological features, and chronological age is the target variable. 

More recent studies have also explored deep neural networks to predict brain age using raw neuroimaging data \cite{cole2017predicting, jonsson2019brain, peng2021accurate, jiang2020predicting, shah2022mri, shah2023multi, bashyam2020mri} and results demonstrate that deep neural networks outperform traditional machine learning approaches given sufficient training data \cite{bashyam2020mri, cole2017predicting, ito2018performance}. Since deep learning methods perform automatic feature extraction from raw structural MRI data, it allows capturing previously unseen imaging signatures related to aging in the brain and makes the model less prone to any biases from pre-processing steps, making it more generalizable.

\subsection{Systematic Bias in Predicted Brain Age}
\label{sec:systematicbias_literature}

In brain age prediction, predicted biological age is often observed to be systematically biased towards the cohort's mean age \cite{liang2019investigating, le2018nonlinear, smith2019estimation, treder2021correlation, beheshti2019bias} affected by regression to the mean (RTM) effect, limiting its potential clinical utility. This causes an unexpected overestimation of predicted brain age in young subjects and underestimation among old subjects. Historically, the RTM effect has been attributed to within-subject and between-subject variability \cite{gardner1973some}. This systematic bias in predicted brain age is not specific to the choice of learning algorithm, data sample imbalance across age groups, or imaging data heterogeneity due to different scanners \cite{liang2019investigating}. Since brain age prediction is traditionally formulated as a regression problem, RTM is a characteristic phenomenon of regression analysis. 

Studies that aim to mitigate this systematic bias propose post-hoc correction methods where predicted age is scaled by slope and intercept derived from regression of predicted age or BrainAGE \cite{cole2018brain, de2019population, beheshti2019bias} on chronological age. Le \etal~\cite{le2018nonlinear} used chronological age as a covariate when analyzing group-level differences in BrainAGE, whereas Cole \etal~\cite{cole2018brain} did not include chronological age in the final adjustment scheme. However, it increased the variance in predicted BrainAGE \cite{beheshti2019bias}. Other studies \cite{beheshti2019bias, de2019population} included chronological age in the final age adjustment, but these methods are likely to be inaccurate when the age range of the independent testing dataset differs from the age range of the model's training data. Recently, Zhang \etal~\cite{zhang2023age} found that these correction methods do not properly address the systematic bias in predictions. Experiments from that study also show that even though linear \cite{beheshti2019bias, cole2018brain} and quadratic \cite{smith2019estimation} correction methods push average BrainAGE close to zero, bias in BrainAGE for same-age subjects gets worse.

More fundamentally, correcting the predicted BrainAGE in a two-step process by explicitly controlling for age would make downstream analysis questionable. This highlights the need to develop a direct method that addresses systematic bias in brain age prediction and is more accurate in predicting normal aging. 

\subsection{Regression as Ordinal Classification}
\label{sec:regascls_literature}

Predicting brain age from imaging data is an ordinal classification task (also known as ordinal regression) since the labels exhibit a natural order. \etal~\cite{gutierrez2015ordinal} conducted a comprehensive exploration of ordinal classification methodologies, categorizing them into three main groups: naive approaches using regression or nominal classification methods, ordinal binary decomposition, and threshold models. However, the efficacy of ordinal decomposition approaches relies heavily on task-specific decomposition strategies, while threshold models demand meticulous calibration of hyperparameters to achieve optimal convergence \cite{rosati2022novel}. In this study, we compare our approach with the nominal classification and regression techniques previously documented in the literature. 

In computer vision, it is shown that classification can outperform regression in many tasks, such as age estimation from face images \cite{pan2018mean, lanitis2004comparing, rothe2018deep}, object counting \cite{liu2019counting}, and depth estimation \cite{cao2017estimating}. The target space is discretized into same-size intervals, and surprisingly, models are more accurate in predicting a range of values rather than estimating actual values on a continuous scale. The exact reason for classification outperforming regression has been less explored before. Zhang \etal~\cite{zhang2023improving} suggest that classification benefits from its ability to learn high entropy feature representations compared to regression, which accounts for the performance gap. Inspired by these insights, we transform the task of brain age prediction from regression to multi-class classification. In brain age prediction, the target output follows a continuous scale consisting of the human life age span. Despite the performance improvement, classification models treat each class label independently from each other, where each wrong prediction is penalized equally. For instance, given a sample with a true age of $53$, cross-entropy (CE) penalizes the model by the same magnitude if the wrong prediction was $21$ or $52$. Hence, the ordinal relationship between target labels is not accurately captured in learned representations of brain age using CE or other loss functions proposed in previous studies \cite{pan2018mean, zhang2023improving} (~\sectionautorefname~\ref{sec:exp_and_results}).

One of the initial works that proposed deep learning-based classification for age estimation from facial images was by Rothe \etal~\cite{rothe2018deep}, where they used the expected mean of softmax weights as the estimated age. Pan \etal~\cite{pan2018mean} also used softmax expected value for age estimation with an additional mean-variance loss used in training. Mean loss minimizes the difference between the mean of the estimated distribution and the ground truth, while the variance loss minimizes the variance of the estimated distribution, resulting in a concentrated distribution. Different from these approaches, Zhang \etal~\cite{zhang2023improving} observed that classification allows learning high entropy feature representation with a more diverse feature set compared to regression. They introduce an Euclidean distance-based loss with mean squared error (MSE) loss for regression to increase the marginal entropy such that learned features are spread out while preserving target ordinality. The latter two studies \cite{pan2018mean, zhang2023improving} also highlight preserved ordinality in learned feature space from their proposed approaches. However, for brain age prediction, results show that this is not the case when compared to a regression model (~\sectionautorefname{~\ref{sec:exp_and_results}}).

\section{Methods}
\label{sec:methods}

\begin{figure*}[t!]
    \centering
    \includegraphics[trim={0 0.1cm 0 0.8cm},clip,width=\linewidth]{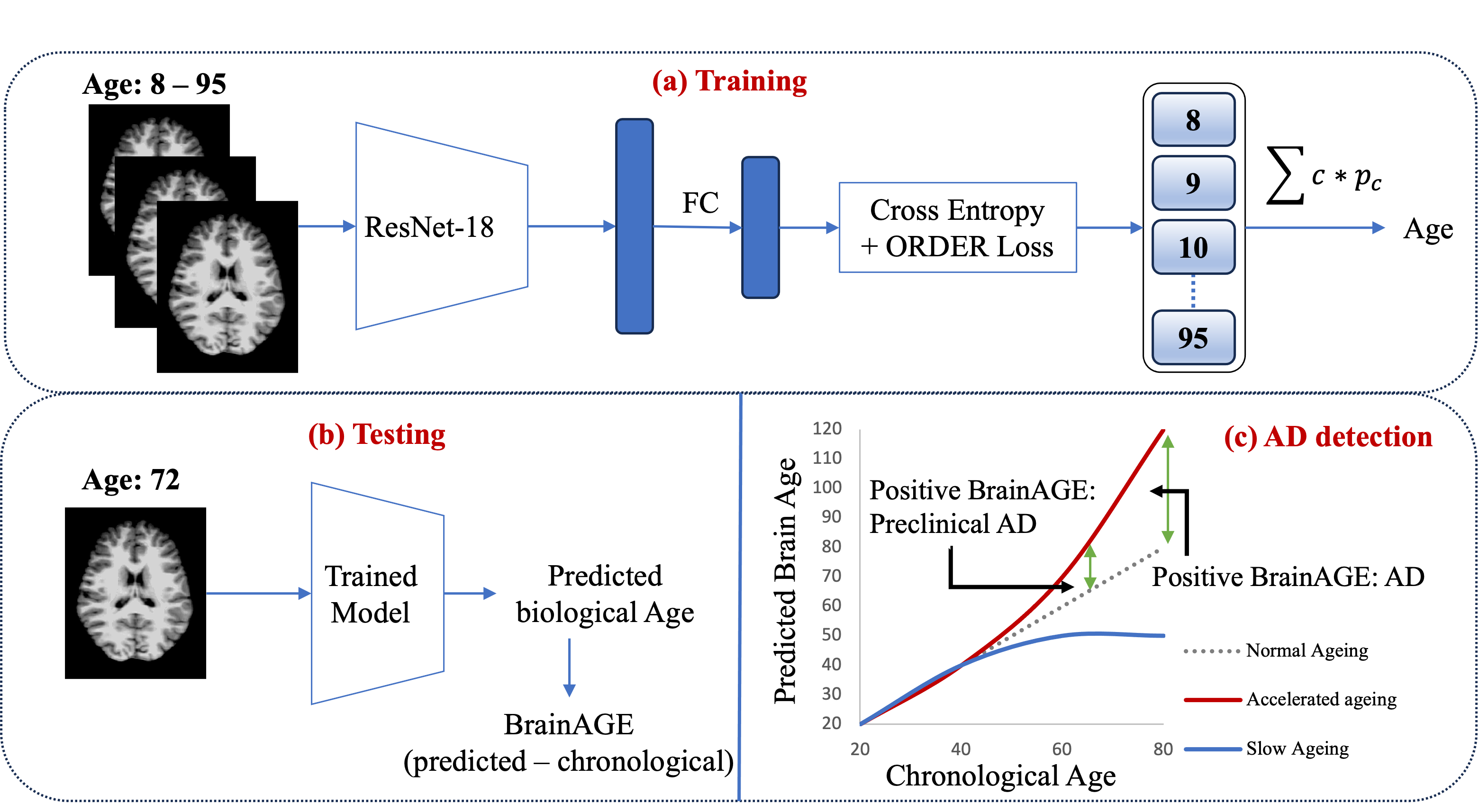}
    \caption{Overview of proposed brain age prediction framework. (a) A 3D ResNet-18 model is trained using lifespan cohort with cross entropy and ORDER losses. Age is calculated as the weighted average of class probabilities from the softmax classifier. (b) At inference, the Brain Age Gap Estimate (BrainAGE) is calculated as the difference between predicted biological age and actual chronological age. (c) The trajectory plot offers a visual interpretation of predicted BrainAGE and its associations with aging patterns. The preclinical AD stage is when the patient behaves cognitively normal, but underlying changes in the brain due to accelerated aging happening at a subtle rate can be captured using BrainAGE.}
    \label{fig:framework_overview}
\end{figure*}

\figureautorefname{\ref{fig:framework_overview}} gives an overview of our framework for robust brain age prediction. In this section, we first describe our proposed ORDER loss that encodes ordinal information within target labels into learned feature space. Then, in addition to MAE, we define two metrics to measure our model's performance in preserving ordinality and minimizing systematic bias compared to established methods.

\subsection{ORDER Loss}
\label{sec:ordinal_distance_regularization_loss}

To better understand the intuition behind the proposed ORDER loss, we first review the original cross-entropy loss ($L_{CE}$), which is formulated as:

\vspace*{-5mm}
\begin{equation}
\label{eq:ce_loss1}
    \begin{split}
    {L}_{CE} & = -\frac{1}{N}\sum_{i=1}^{N}\log (\hat{y}_{i}) \\
    & = -\frac{1}{N}\sum_{i=1}^{N}\log \frac{e^{W^T_{{y}_{i}}{x}_{i}}}{\sum_{j=1}^{C} e^{W^T_{{y}_{j}}{x}_{i}}}
    \end{split}
\end{equation}
where $x_i$ is input to the last fully connected layer corresponding to $i$-th sample from training data $N$, $y_i$ is the hot encoding of the true label, $\hat{y}_i$ is the predicted probability, and $W^T_{{y}_{j}}$ is $j$-th column of last fully connected layer ($j \in [1, C], C$ is number of classes). $W^T_{{y}_{i}}{x_i}$ often denoted as ${z_i}$, is the target logit of $i$-th sample \cite{pereyra2017regularizing}.

\vspace*{-3mm}
\begin{equation}
\label{eq:ce_loss2}
    \begin{split}
    {L}_{CE} & = -\frac{1}{N}\sum_{i=1}^{N}\log \frac{e^{z_i}}{\sum_{j=1}^{C} e^{z_j}}
    \end{split}
\end{equation}

In brain age prediction, our main aim is to understand how the dependent variable (age) changes with variations in independent variables (imaging features). Given a sample from class $i$, cross-entropy loss forces $z_i > z_j (\forall j\neq i)$. However, when the class labels are ordered, it does not guarantee that learned feature representation follows the same order, i.e., $z_{i} < z_{i+1} < z_{i+2} <... < z_{C}$ and $z_{i} > z_{i-1} > z_{i-2} >... > z_{1}$. Even though $L_{CE}$ increases the marginal entropy of feature space, resulting in a diverse feature set, the marginal ordering between class labels is not correctly captured. Keeping the diversity of features from $L_{CE}$ intact, we adjust the target logit $z_i$ with corresponding feature vector $x_i$'s distance to other features $x_j (\forall j\neq i)$ in a batch of samples, weighted by distance between class labels. 

\vspace*{-7mm}
\begin{equation}
    \label{eq:new_target_logit}
    \begin{split}
        z' = W^T_{{y}_{i}}{x_i} + \varphi(x_i)
    \end{split}
\end{equation}
where,

\vspace*{-7mm}
\begin{equation}
\label{eq:ordinal_distance_encoded_regularization_loss_term}
    \begin{split}
        \varphi(x_i) = \frac{1}{N-1}\sum_{j=1, i\neq j}^{N}|i - j||\Bar{x}_i - \Bar{x}_j|_{manh}
    \end{split}
\end{equation}
$\Bar{x}$ is $L_2$ normalized vector $\Bar{x} = x/max(||x||_2)$. Substituting \equationautorefname\ref{eq:new_target_logit} in \equationautorefname\ref{eq:ce_loss2} we get new loss $L_T$, which can be decomposed into $L_{CE}$ and ORDER loss $(L_{ORDER})$ 

\vspace*{-3mm}
\begin{equation}
\label{eq:our_loss_total}
    \begin{split}
    {L}_{T} & = -\frac{1}{N}\sum_{i=1}^{N}\log \frac{e^{W^T_{{y}_{i}}{x}_{i}+\varphi(x_i)}}{\sum_{j=1}^{C} e^{W^T_{{y}_{j}}{x}_{i}}} \\
    & = -\frac{1}{N}[\sum_{i=1}^{N}\log \frac{e^{W^T_{{y}_{i}}{x}_{i}}}{\sum_{j=1}^{C} e^{W^T_{{y}_{j}}{x}_{i}}} + \sum_{i=1}^{N}\varphi(x_i)] \\
    & = -\frac{1}{N}\sum_{i=1}^{N}\log \frac{e^{W^T_{{y}_{i}}{x}_{i}}}{\sum_{j=1}^{C} e^{W^T_{{y}_{j}}{x}_{i}}} \\
    & -\frac{1}{N(N-1)}\sum_{j=1, i\neq j}^{N}|i - j||\Bar{x}_i - \Bar{x}_j|_{manh} \\
    & = L_{CE} + L_{ORDER}
    \end{split}
\end{equation}

We use the Manhattan distance to calculate the distance between two features $x_i$ and $x_j$ in high-dimensional space. Euclidean distance is the most common metric to measure similarity or distances between two data points. However, Aggarwal \etal~\cite{aggarwal2001surprising} found that, due to the curse of dimensionality in high-dimensional space, the sparsity of features is significantly high, making them almost equidistant from each other. The ratio between the closest and farthest points from a reference sample approaches $1$ in high-dimension space \cite{domingos2012few}. This further explains the inability of a classification model to capture ordinal information. We explored different orders of distance metrics for ORDER loss, but Manhattan distance performed best (see \sectionautorefname\ref{sec:ablation}). 

\subsection{Evaluation Metrics}

\subsubsection{Measuring Ordinality}
\label{sec:ordinality_metric}

To the best of our knowledge, there are no defined metrics in the literature that measure the ordinality of feature representations from a deep learning model with reference to the order of ground truth. Given $n$ images and $c$ ordered classes, we first obtain $n$ features of $512$ dimensions from the penultimate layer of a trained model ${\{x_1, x_2, ..., x_n\}}$. From those features, we calculate $c$ feature centroids ${\{f_1, f_2, ..., f_c\}}$ using ground-truth labels corresponding to each class. After that, Manhattan distances between $f_1$ and other feature centroids can be calculated as $D = \{d_{12}, d_{13}, ... d_{1c}\}$ where,

\vspace*{-3mm}
\begin{equation}
\label{eq:feature_centroid_distance}
    \begin{split}
    d_{ij} = |d_i - d_j|_{manh}
    \end{split}
\end{equation}

Since class labels here are age values in a chronologically increasing order, we get $C = {\{1, 2, ..., (c-1)\}}$ as the distance of the first class to others. We define the ordinality metric as the Pearson correlation coefficient between $D$ and $C$. For a model that perfectly captures ordinal relationships in feature representations, the ordinality score is close to $1$. Pearson correlation between two continuous variables measures how much change in one variable is associated with a proportional change in the other variable. Using this metric, our model's performance in capturing age-related order information from labels compared to other approaches is evaluated (see \tableautorefname \ref{tab: ordinal_bias_results}). An ordinality score close to +$1$ indicates that the learned features have a similar ranking order as their corresponding ground-truth labels and a lower value indicates otherwise.

\vspace*{-3mm}
\subsubsection{Quantifying Systematic Bias}
\label{sec:quantifying_systematicbias}

Previous approaches discussed in \sectionautorefname\ref{sec:quantifying_systematicbias} that propose post-hoc correction methods use correlation of predicted BrainAGE and chronological age as a measure of underlying systematic bias \cite{le2018nonlinear, liang2019investigating}. Using chronological age to adjust BrainAGE would reduce age dependence on BrainAGE, i.e., $r = 0$. However, it does not address the inherent systematic bias effect caused due to regression. Additionally, this correction method would be questionable when the test dataset does not have the same age range as the training dataset. 

To objectively quantify systematic bias caused by regression to the mean effect, we compare the predicted BrainAGE at one standard deviation away from mean \cite{gardner1973some}, i.e., for values less than $(\mu - \sigma)$ and greater than $(\mu + \sigma)$, where $\mu$ and $\sigma$ are mean and standard deviation of target age values of the test set. We refer to these two groups as Systematic Bias - Left and Right (SB-L, SB-R). Since there is an overestimation of predicted biological age in young subjects and an underestimation in old subjects, the bias causes higher BrainAGE and lower BrainAGE values for those respective sub-groups. These scores are compared for different methods, and a value closer to $0$ indicates better performance in addressing systematic bias (see \tableautorefname{\ref{tab: ordinal_bias_results}}).

\section{Experiments and Results}
\label{sec:exp_and_results}

We evaluate our proposed brain age prediction framework and other baseline methods on a combined healthy cohort using three different metrics specific to this task. Evaluation metrics include MAE, Ordinality, and Systematic Bias scores. 

A 3D ResNet-18 was adopted as the base deep-learning model, and input to the model are 3-dimensional MRI scans with a batch size of $4$. Stratified oversampling was employed in classification models, and for regression, samples were stratified based on age groups $(8-12, 12-16, ..., 92-96)$ to perform oversampling. We used AdamW optimizer with a $1e^{-3}$ learning rate and weight decay of $1e^{-2}$. Each model was trained for $100$ epochs and with early stopping to avoid over-fitting. All experiments were performed on NVIDIA's $A100 \ 80GB$ GPUs to train, validate, and test the models. 

\subsection{Datasets and Preprocessing}
\label{sec:data_and_preprocessing}

Since most medical imaging datasets are part of multi-center studies, differences in scanners, imaging protocols, variations in vendors, and their hardware account for heterogeneity in data. Deep learning models are known to be robust against heterogeneity in data. In fact, including more heterogeneous data in model training improves its generalization on out-of-distribution data \cite{maartensson2020reliability}. With that consideration, a combined lifespan cohort of $7,377$ T1-weighted MRI scans of healthy participants collected from five different public sources was used in model training.


\textbf{Lifespan cohort}: All the age prediction models were trained, validated, and tested on a healthy cohort (age: 8-95 years) collected from $(1)$ NACC Uniform Data Set (UDS) from 1999 to March 2021 $(2)$ OASIS $(3)$ ICBM $(4)$ IXI (\href{http://brain-development.org/ixidataset/}{http://brain-development.org/ixidataset/}) and $(5)$ ABIDE. These cohorts included both 1.5T and 3T scans with predominantly Caucasian participants but also included other race/ethnic groups. The number of samples and age range per cohort are summarized in \tableautorefname{\ref{tab: datasets}}. 

All five cohorts were preprocessed using an in-house data preprocessing pipeline. T1-weighted MR images were first aligned to the MNI template with rigid transformation, and then intensity normalized and conformed using FreeSurfer v7 to generate preprocessed images at 1 mm isotropic voxels with a 256 x 256 x 256 matrix.

\textbf{Discovery cohort}: Additionally, $1,584$ MRI scans were collected from the Alzheimer's Disease Neuroimaging Initiative (ADNI; \href{https://adni.loni.usc.edu/}{https://adni.loni.usc.edu/}) database containing a mix of healthy, cognitively impaired, and AD patients. This cohort was used as an independent testing and discovery dataset to evaluate model performance in predicting age and its ability to differentiate clinical groups in AD. Priority was given to scans with matching PET data and participants who had longitudinal follow-ups. For healthy controls (HCs), a random subset was selected from the overall ADNI set and included in this analysis. The diagnostic status was determined based on ADNI clinical data. In this analysis, HC (N=678) participants had normal cognition and did not convert to MCI or AD in follow-up visits. HC to MCI converters (HC-MCI, N=179) are participants who had normal cognition at baseline but converted to MCI during follow-up. MCI-stable (MCIs, N=432) participants had a baseline diagnosis of MCI and stayed unchanged in follow-ups. MCI to AD converters (MCI-AD, N=139) are those participants with MCI diagnosis at baseline and subsequently converted to AD. AD (N=156) patients are those who were diagnosed with AD at baseline.  

\begin{table}[h!]
\centering
\begin{tabular}{|>{\centering\arraybackslash}p{1.1cm}|>{\centering\arraybackslash}p{1cm}|>{\centering\arraybackslash}p{2.4cm}|>{\centering\arraybackslash}p{2cm}|} 
 \hline
 \textbf{Dataset} & \textbf{Count} & \textbf{Age Range} (yrs) & \textbf{Mean $\pm$ STD}\\  
 \hline
 NACC & 4,132 & 18 - 95 & 67.5 $\pm$ 10.8 \\ 
 \hline
 OASIS & 1,432 & 8 - 94 & 27.9 $\pm$ 20.7 \\
 \hline
 ICBM & 1,101 & 18 - 80 & 37.6 $\pm$ 15.4 \\
 \hline
 IXI & 536 & 20 - 86 & 48.8 $\pm$ 16.5 \\
 \hline
 ABIDE & 176 & 18 - 56 & 26.1 $\pm$ 7.0 \\
 \hline
 ADNI & 1,584 & 55 - 98 & 73.3 $\pm$ 7.3\\
 \hline
\end{tabular}
\caption{Age range with distribution and number of samples for each cohort. The lifespan cohort comprises NACC, OASIS, ICBM, IXI, and ABIDE, whereas the Discovery cohort consists of samples from the ADNI cohort.}
\label{tab: datasets}
\end{table}

\vspace*{-5mm}
\subsection{BrainAGE prediction}
\label{sec:mae_results}

We compare our proposed method's performance in predicting the brain age of healthy individuals from lifespan cohort to four baseline methods, including two regression and two classification models (\tableautorefname\ref{tab: mae_results}). For classification models, age values were rounded off to the closest integer and assigned respective class labels. Only $535 (7.3\%)$ samples from the lifespan cohort had non-integer age values. 

\begin{figure*}[t!]
    \centering
    \includegraphics[trim={0 0.4cm 0 0},clip,width=\linewidth]{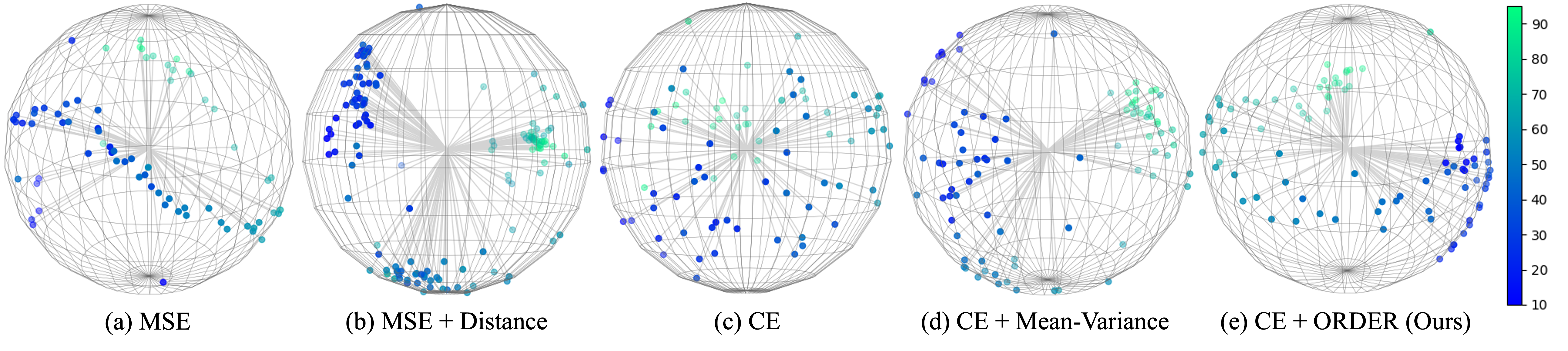}
    \caption{t-SNE visualization of embeddings from models' penultimate layer: (a) When using MSE loss, embeddings maintain ordinal relationships but are tightly packed, resulting in a low-entropy feature space (b) MSE with Euclidean distance loss spreads out embeddings but struggles to preserve ordinal relationships accurately (c) Cross-entropy (CE) further spreads embeddings, creating a high-entropy space, but at the cost of losing ordinal information (d) Mean-variance loss combined with cross-entropy creates a high-entropy feature space and slightly improves ordinality (\tableautorefname{\ref{tab: ordinal_bias_results}}). (e) ORDER loss combined with cross-entropy achieves the best balance: it accurately preserves ordinality, maintains a high-entropy space, and improves overall performance. Embeddings are colored-coded based on their ground truth age values $[10-95]$.}
    \label{fig:embeddings_plot}
\end{figure*}

Our model performed best on the healthy test set with MAE $2.56$, outperforming standard MSE and cross-entropy loss models. Among other competing methods, the classification model with mean-variance loss performed best. The model with cross-entropy loss outperforms the MSE model due to its ability to learn high entropy features (\figureautorefname{\ref{fig:embeddings_plot}}), where inter-class features are spread out, and intra-class features are compact \cite{boudiaf2020unifying}. Surprisingly, adding Euclidean distance-based regularizer to MSE loss did not improve the regression model's performance. Our method's performance is also significantly better than MAE reported by prior studies using regression analysis \cite{ito2018performance, jonsson2019brain, cole2017predicting}, however, on different cohorts.

\begin{table}[h!]
\centering
\begin{tabular}{|>{\centering\arraybackslash}p{5cm}|c|} 
 \hline
 \textbf{Method} & \textbf{MAE} \\ 
 \hline
 MSE & 3.93 \\ 
 \hline
 MSE + Distance \cite{zhang2023improving} & 4.57 \\
 \hline
 CE \cite{rothe2018deep} & 3.33 \\
 \hline
 CE + Mean-Variance \cite{pan2018mean} & \underline{2.65} \\
 \hline
 CE + ORDER (Ours) & \textbf{2.56} \\ 
 \hline
\end{tabular}
\caption{Brain age prediction results on lifespan cohort. MAE measures the difference between predicted and actual chronological age on the same test set. \textbf{Bold} numbers represent the best results, while \underline{underlined} numbers represent second-best results.}
\label{tab: mae_results}
\end{table}

\vspace*{-2mm}
\subsection{Ordinality and Systematic Bias}
\label{sec:ordinal_systematicbias_results}

We further evaluate our model's ability to preserve ordinality and address systematic bias in predicted BrainAGE using metrics defined in \sectionautorefname\ref{sec:ordinality_metric} and \sectionautorefname\ref{sec:quantifying_systematicbias}. As expected, the model with MSE loss had the highest ordinality score (\tableautorefname\ref{tab: ordinal_bias_results}). Our classification model with ORDER loss had an ordinality score much closer to standard MSE loss than other methods, demonstrating its effectiveness in learning ordinal information. \figureautorefname{\ref{fig:embeddings_plot}} offers a visual comparison of learned feature space using different loss functions to confirm this further. 

Furthermore, the model with ORDER loss also performed best in reducing systematic bias measured by average BrainAGE values at one standard deviation away from the mean. The mean of the test set was $53.4$ with a standard deviation of $22.2$. Hence, the bias scores reported in \tableautorefname\ref{tab: ordinal_bias_results} are BrainAGE values for age $<31.2$ (SB-L) and age $>75.6$ (SB-R). Values closer to zero reflect a better reduction in systematic bias. Both MSE-based models had a higher systematic bias due to the inherent RTM effect. Due to its ability to learn class-specific and diverse feature sets, cross-entropy loss reduces bias effects for SB-L and SB-R groups. Incorporating order information allows the model to learn the relative ranking of labels, further improving ordinal classification performance.

\begin{table}[h!]
\centering
\begin{tabular}{|>{\centering\arraybackslash}p{3cm}|c|c|c|}
\hline
\multirow{2}{*}{\textbf{Method}} & \multirow{2}{*}{\textbf{Ordinality}} & \multicolumn{2}{l|}{\textbf{Systematic Bias}} \\ \cline{3-4} 
  &  &  \textbf{SB-L} & \textbf{SB-R} \\ \hline
 MSE & \textbf{0.99} & 3.4 & -4.2 \\ 
 \hline
 MSE + Distance & 0.95 & 4.8 & -4.1 \\
 \hline
 CE & 0.31 & 1.1 & -3.6 \\
 \hline
 CE + Mean-Variance & 0.58 & \underline{0.4} & \underline{-4.2} \\
 \hline
 CE + ORDER & \underline{0.98} & \textbf{0.1} & \textbf{-2.5} \\
 \hline
\end{tabular}
\caption{Performance evaluation of all methods in preserving ordinality and addressing systematic bias in brain age prediction using metrics defined in \sectionautorefname\ref{sec:ordinality_metric} and \sectionautorefname\ref{sec:quantifying_systematicbias}. }


\label{tab: ordinal_bias_results}
\end{table}

\subsection{Alzheimer's Disease detection}


AD has a prolonged preclinical phase where brain changes manifest subtly as accelerated aging \cite{long2019alzheimer}. \figureautorefname{\ref{fig:framework_overview}} illustrates this phase, showing accelerated aging diverging slightly from normal aging. MCI, a pre-dementia stage, involves greater cognitive decline than typical aging \cite{selkoe1997alzheimer}. BrainAGE can help detect and monitor this stage early.

The discovery cohort (\sectionautorefname\ref{sec:data_and_preprocessing}) obtained from ADNI with five clinical groups was used to test BrainAGE prediction using different methods. Trained models were applied to this cohort using the abovementioned methods to calculate BrainAGE. These five groups were ranked $[1-5]$ in an increasing order of disease severity as HC $<$ HC-MCI $<$ MCI-stable $<$ MCI-AD $<$ AD. Since disease severity is proportional to accelerated aging, we expect the average predicted BrainAGE to follow the same order. Pearson correlation is calculated between the model's predicted BrainAGE and rank of disease severity. A high correlation would indicate the model's ability to accurately characterize aging signatures along the AD continuum via estimated BrainAGE. From \tableautorefname{\ref{tab: ad_results}}, we see that only the model with MSE and our proposed loss have a high correlation. 

\begin{figure}[!htp]
    \centering
    \includegraphics[trim={0 0.5cm 0 0.1cm},clip,width=\linewidth]{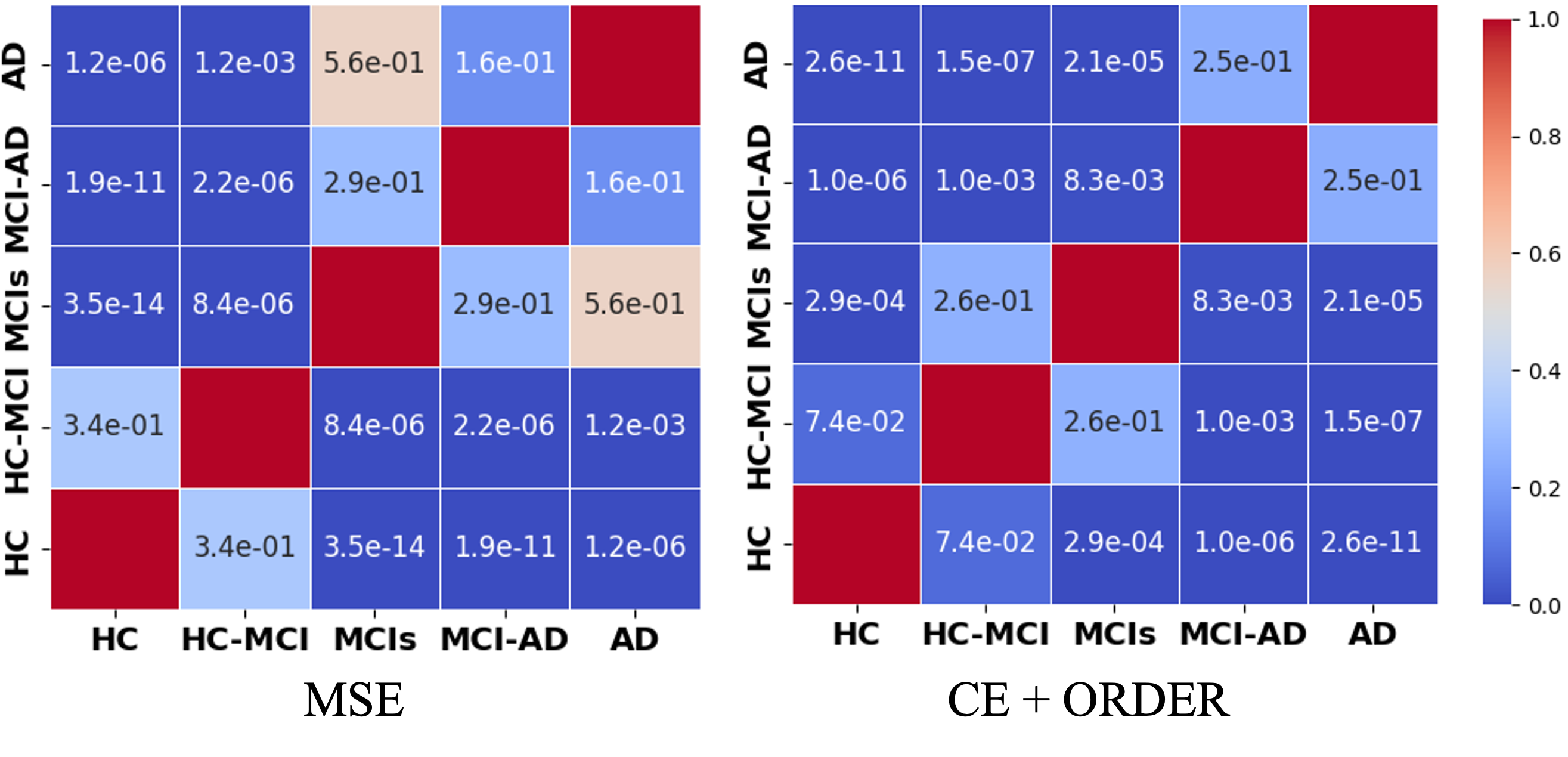}
    \caption{Heatmap of statistical significances between the five clinical groups of AD calculated as $p$ values from a t-test on predicted BrainAGE from respective groups, for MSE and cross-entropy with ORDER loss models.}
    \label{fig:significance_heatmap}
\end{figure}

We further compare the ability of MSE and ORDER loss models to detect subtle differences between these clinical groups accurately. \figureautorefname{\ref{fig:significance_heatmap}} shows the MSE model had a more disruptive trend in predicted BrainAGEs between groups, i.e., there was a higher difference between AD and MCI-AD $(p=0.16)$ compared to AD and MCI-stable $(p=0.56)$. Whereas the ORDER loss model had an overall consistent trend in statistical significance between groups associated with actual disease severity, highlighting its better discriminative power. It was also able to better detect differences between HC and HC-MCI subjects $(p=0.07)$ compared to MSE $(p=0.34)$, which is crucial for early AD detection. Although our model's performance wasn't as strong as MSE in distinguishing between HC-MCI and MCI-stable, we posit that this could be attributed to the definitions of clinical groups used here. The absence of clinical tools to definitively differentiate HC-MCI from MCI-stable groups, given that subjects exhibit normal cognitive behavior and no discernible symptoms despite age-related brain changes, might contribute to this outcome. We plan to work with clinicians to further investigate these observations from both groups.



\vspace*{-3mm}
\begin{table}[h!]
\centering
\begin{tabular}{|>{\centering\arraybackslash}p{1.3cm}|>{\centering\arraybackslash}p{0.6cm}|>{\centering\arraybackslash}p{0.6cm}|>{\centering\arraybackslash}p{0.7cm}|>{\centering\arraybackslash}p{0.6cm}|>{\centering\arraybackslash}p{0.6cm}|>{\centering\arraybackslash}p{0.8cm}|} 
 \hline
 \textbf{Method} & \textbf{HC} & \textbf{HC-MCI} & \textbf{MCIs} & \textbf{MCI-AD} & \textbf{AD} & \textbf{Corr.}\\
 \hline
 MSE & -1.2 & -0.8 & -0.3 & 0.8 & 1.5 & 0.98\\ 
 \hline
 {MSE + \par Distance} & -2.7 & -1.9 & -1.7 & -0.9 & 0.9 & 0.94\\
 \hline
 CE & -1.9 & -1.5 & -3.4 & -2.3 & -4.1 & -0.75 \\
 \hline
 CE + MV & -1.6 & -0.3 & -0.5 & 0.8 & 2.8 & 0.94\\
 \hline
 {CE + \par ORDER} & -1.5 & -0.8 & -0.3 & 1.2 & 2.0 & 0.98\\
 \hline
\end{tabular}
\caption{Average BrainAGE values across the five clinical groups of AD. Last column is the Pearson correlation between average BrainAGE values and disease severity of clinical groups in increasing order from HC to AD. MV: Mean-Variance.}
\label{tab: ad_results}
\end{table}

\subsection{Ablation studies}
\label{sec:ablation}

\noindent\textbf{Distance Metric}: We explored different $L_k$ norm distance metrics for ORDER loss and found Manhattan distance best performing across all evaluation measures. $L_k$ norm distance between two points $x$ and $y$ in high-dimensional space, given $(x, y \in R^{d})$, is be defined as:

\vspace*{-5mm}
\begin{equation}
\label{eq:lk_distance_function}
    \begin{split}
    {L}_{k}(x, y) = \sum_{i=1}^{d}[||x^i - y^i||^{k}]^{\frac{1}{k}} 
    \end{split}
\end{equation}

Aggarwal \etal~\cite{aggarwal2001surprising} showed that Manhattan distance $(k=1)$ is a more suitable distance metric than Euclidean $(k=2)$ for high-dimensional data. They recommended using $k\leq1$ to improve downstream classification performance. Later studies showed that fractional distance metrics, i.e., $(k<1)$, do not systematically address the issue of the curse of dimensionality \cite{mirkes2020fractional} but should be a choice depending on the training data distribution. For the high-dimensional neuroimaging dataset used here, results indicate that Manhattan distance is more accurate in preserving ordinality and improving class separability compared to Euclidean or fractional distance metrics (see \tableautorefname{\ref{tab: lk_distance_results}}). 

\vspace{5pt}
\noindent\textbf{ORDER loss with Classification vs. Regression}: We experimented with the proposed ORDER loss using both classification and regression frameworks. As discussed in the paragraph above, since Euclidean and Manhattan distances performed significantly better than fractional distances, we explored regression models with our loss for $k=\{1, 2\}$ (\tableautorefname{\ref{tab: lk_distance_results}}). Results show that distance-based regularization does not work well in regression models. Our model with cross-entropy loss and Manhattan distance-based ordinal regularization performed best across the three metrics. 

\begin{table}[h!]
\centering
\begin{tabular}{|c|c|c|c|c|c|}
\hline
\multirow{2}{*}{\textbf{$k$}} & \multirow{2}{*}{\textbf{Loss}} & \multirow{2}{*}{\textbf{MAE}} & \multirow{2}{*}{\textbf{Ordinality}} & \multicolumn{2}{l|}{\textbf{Systematic Bias}} \\ \cline{5-6} 
  &  &  &  & \textbf{SB-L} & \textbf{SB-R} \\ \hline
 $1$/$2$ & CE & 6.05 & 0.85 & 5.31 & -5.19 \\ 
 \hline
 $2$/$3$ & CE & 18.51 & 0.13 & 30.67 & -28.27 \\
 \hline
 $1$ & \textbf{CE} & \textbf{2.56} & \textbf{0.98} & \textbf{0.11} & \textbf{-2.5} \\
 \hline
 $1$ & MSE & 4.66 & \underline{0.95} & 2.19 & -4.98 \\
 \hline
 $2$ & CE & \underline{2.90} & 0.10 & \underline{0.93} & \underline{-3.04} \\
 \hline
 $2$ & MSE & 4.57 & \underline{0.95} & 4.83 &  -4.13 \\
 \hline
\end{tabular}
\caption{Ablation studies on the proposed framework components evaluated by MAE, ordinality, and systematic bias scores. $k$ denotes different $L_k$-norm distance metrics defined in \equationautorefname{\ref{eq:lk_distance_function}}}.
\label{tab: lk_distance_results}
\end{table}

\vspace*{-7mm}
\section{Conclusion}
\label{sec:conclusion}

This paper proposes a novel ordinal-distance regularization loss for robust brain age prediction using deep learning. ORDER loss in an ordinal classification framework outperforms regression-based brain age prediction methods, reduces systematic bias in predictions, and preserves ordinality in learned feature space. Improved performance is attributed to ordering information encoded in the model using ORDER loss and the ability of cross entropy loss to learn high entropy feature representations. The predicted BrainAGE from this model is a more reliable imaging biomarker for diagnosing AD and predicting its early onset. We believe this framework can be generalized to other regression tasks to improve prediction and address the RTM effect if present, which we aim to investigate further in future work.

\vspace{5pt}
\noindent\textbf{Acknowledgments.} This research received support from the National Institute on Aging of the National Institutes of Health under Award Numbers R01AG069453,  P30AG072980,  RF1AG073424, Banner Alzheimer’s Foundation and the Arizona Department of Health Services to Arizona Alzheimer’s Research Center. We thank Arizona State University Research Computing (ASURC) for hosting and maintaining our computing resources.

\newpage

{\small
\bibliographystyle{ieee_fullname}
\bibliography{egbib}
}

\end{document}